\documentclass[useAMS,usenatbib,referee]{mn2e}
\usepackage{graphicx,amsmath,times}

\title[The chemical evolution as a function of metallicity]{Modelling the chemical evolution
of molecular clouds as a function of metallicity}
\author[E. M. Penteado, H. M. Cuppen and H. J. Rocha-Pinto]{E. M. Penteado$^{1}$\thanks{E-mail:
e.monfardini@science.ru.nl}, H. M. Cuppen$^{1}$ and H. J. Rocha-Pinto$^{2}$\\
$^{1}$Institute for Molecules and Materials, Radboud University Nijmegen
              Heyendaalsweg 135, 6525 AJ Nijmegen, The Netherlands\\
$^{2}$UFRJ, Observat\'orio do Valongo, Ladeira Pedro Ant\^onio 43, 20080-090, Rio de Janeiro, Brazil}
\begin{document}


\pagerange{\pageref{firstpage}--\pageref{lastpage}} 

\maketitle

\label{firstpage}

\begin{abstract}
The Galaxy is in continuous elemental evolution. Since new elements produced by dying stars are delivered to the 
interstellar medium, the formation of new generations of stars and planetary systems is influenced by this metal enrichment. 
We aim to study the role of the metallicity on the gas phase chemistry of the interstellar medium.
Using a system of coupled-ordinary differential equations to model the chemical reactions, we simulate the evolution of the abundance 
of molecules in the gas phase for different initial interstellar elemental compositions. These varying initial elemental compositions 
consider the change in the ``elemental abundances" predicted by a self-consistent model of the elemental evolution of the Galaxy. As far as we are aware, this is the first attempt to combine elemental evolution of the Galaxy and chemical evolution of molecular 
clouds. The metallicity was found to have a strong effect on the overall gas phase composition. With decreasing 
metallicity, the number of long carbon chains was found to increase, the times-cale on which small molecular species are 
increases, and the main form of oxygen changed from O and CO to O$_2$. These effects were found to be mainly due to the change in 
electron, H$_3^+$, and atomic oxygen abundance.
\end{abstract}


\section{Introduction}
Molecular clouds are chemically rich environments where over 180 molecular 
species have been observed and these are the sites of star and planet formation \citep{Herbst:2009}. Molecules play a
central role in determining the thermal budget
of astrophysical bodies and provide crucial building blocks during star and
planet formation. The understanding of such environments has always been a challenge and, for decades, astrochemists have 
developed different models to describe the rich and complex chemistry of the interstellar medium (ISM), which might 
consider only reactions occurring in the
gas phase \citep{Herbst:1973} or also chemistry taking place on the surfaces of interstellar dust 
grains \citep{Tielens:1982}. Although it is well known that many molecules are more
efficiently formed on the surface of astrophysical 
ices, gas phase chemistry has an important role on the chemical evolution of dense molecular clouds and its 
study is still necessary to deepen our knowledge about the evolution of the ISM. Models to describe 
gas phase \citep[\emph{e.g.}][]{Wakelam:2012} and grain surface chemistry \citep[\emph{e.g.}][]{Vasyunin:2012} of the ISM are 
continuously elaborated and are improved by comparisons with laboratory results and with observations.

The most commonly used method to model interstellar gas phase chemistry is based on reaction rate equations. In this 
method, a set of ordinary coupled differential equations, in which 
each equation describes the time variation of the concentration of a molecule, is built accordingly to the 
reactions under consideration. This set of differential equations can only be solved numerically and the number of 
equations follows the number of species considered, typically hundreds, which are normally connected by thousands of 
reactions. The final aim of astrochemical models is to understand how the abundances of the molecules change with time 
according to a set of initial abundances and to a set of astrophysical parameters.

Many molecules are expected to be discovered in the near future because of the
development of a new generation of ground-base interferometers, like the Atacama Large Millimeter/submillimeter Array 
(ALMA), forcing the development of new models able to 
describe such complex environment in more details. Moreover, these instruments will allow us to look at the molecular 
complexity beyond our own Galaxy, probably even at higher redshift. Studies of formation of simple molecules at high 
redshift have been developed \citep{Cazaux:2009} and detection of molecules in extragalactic sources has already been reported
\citep[][and references therein]{Pereira-Santaella:2013}.

\begin{figure*}
    \centering
    \includegraphics[width=14cm]{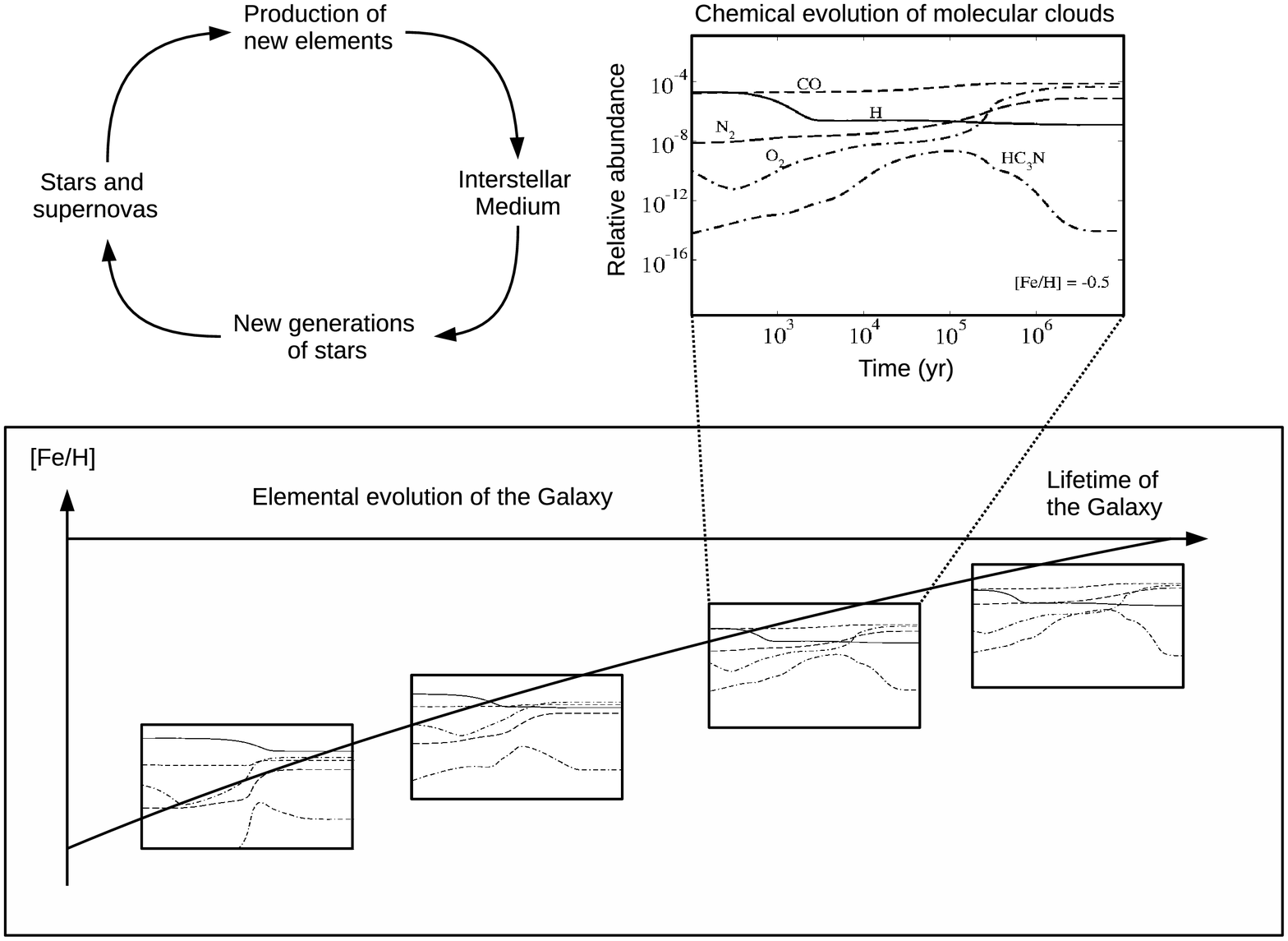}
    \caption{Chemical evolution of molecular clouds connected to the elemental evolution of the Galaxy. Top-left: schematic 
representation of the process of production and delivery of new elements to the ISM. Top-right: an example of chemical 
evolution of a molecular cloud for one specific metallicity. Bottom: representation of the elemental evolution of the Galaxy, 
where molecular clouds evolve with different initial metallicity along the life time (redshift) of the Galaxy.}
       \label{ElemChemEvol}
  \end{figure*}

The present work presents simulations of the variation of abundances of molecules in a cold dense molecular cloud computed by
taking into account the elemental evolution of the Galaxy. \cite{Wakelam:2010B} have also studied the effect of the change in the initial 
elemental abundances and other input parameters but, differently from our study, their choice of the initial abundances 
did not take into account the elemental evolution of the Galaxy. Elements heavier than hydrogen and helium are delivered to the 
ISM during stellar deaths. This process continuously enriches 
molecular clouds so that new generations of stars and planetary systems are born with a different initial elemental 
composition. To this process of element formation and delivery to the ISM, we refer as ``elemental evolution of the 
Galaxy''. To the production and evolution of molecules in dense cold molecular clouds, we refer 
as ``chemical evolution''\footnote{Since this work tracks both elemental and molecular evolution, we decided to use the 
term ``chemical evolution'' to refer only to the evolution of the abundance of molecules in clouds and  
``elemental evolution'' to refer to the production of elements by dying stars. Therefore, the term ``elemental evolution'' 
is a synonymous for the term ``chemical evolution'' used by other authors.}. Figure \ref{ElemChemEvol} shows schematically how the 
production of new elements by dying stars influences the production of molecules deeply inside dense and cold molecular clouds. The 
elemental evolution of the Galaxy can be interpreted as a clock, in which the metallicity, denoted by $[{\rm Fe}/{\rm H}]$, increases
as time passes, so that the Galaxy must be more rich in metals in late than in early times. This can also be thought of in 
terms of redshift, so that galaxies at higher redshift must have lower metallicity, whereas galaxies at lower redshifts 
must have higher metallicity. Therefore, the metallicity represents a chronometer, which describes the accumulation of iron 
in the ISM. \cite{Wheeler:1989} have already noted that the accumulation of iron in the ISM increases monotonically with time.
Therefore, by connecting the elemental evolution of the Galaxy with the chemical evolution of the ISM, the study of the chemistry 
occurring in primordial gas of higher redshift is possible, as well as a comparison to the chemistry of present higher 
metallicity gas. 

As mentioned before, we will limit our study to gas phase chemical models. Ultimately, one would like to repeat this study 
applying a full gas-grain model. However, a pure gas phase chemical model will be a good first step in understanding the 
influence of redshift on the gas phase composition. It is a representative model of quiescent cold clouds. Grains will 
mostly contribute in depleting the gas phase from certain species, but this is more dominant at high densities 
($> 10^5$cm$^{-3}$), and in the formation of new molecules that will enter the gas phase in later stages of the star 
formation sequence where the grains heat up and release their ice mantles. In the present study we will therefore limit 
ourselves to description of the gas phase composition at cold cloud conditions and we aim to understand how the gas phase 
chemistry changes with metallicity. We will further speculate on the possible further reactions that could occur on the 
grain surfaces. One might ask whether it is reasonable to study the galactic chemistry without considering the presence of grains. 
Gas$-$grain models are still in development and many issues still wait to be solved. The present work is the first attempt to combine both 
elemental and molecular evolution, connecting two different formalisms. This brings the necessity of starting with simple considerations, 
like excluding, for now, the presence of grains.

Details of the method and the process to find the initial elemental composition are described in Sec.~\ref{method}, 
while the results and discussion are presented in Sec.~\ref{results}. Finally, Sec.~\ref{conclusions} gives our concluding remarks.

\section{Method}
\label{method}
The applied method consists of three steps: first, the initial abundances of a selection of elements are determined on the 
basis of a standard self-consistent model for the elemental evolution of the Galaxy described by \cite{Timmes:1995}. These abundances
are then depleted to account for the dust and finally given to
a chemical code which calculates the time dependent chemical evolution of the molecular cloud. 

\subsection{Initial elemental abundances}
The initial elemental abundances as a function of metallicity were obtained from a model for the elemental evolution of 
the Galaxy presented by \cite{Timmes:1995}. These authors simulated the production of 76 stable isotopes by Type I and II supernovae,
from hydrogen to zinc, through a dynamical and chemical model for the Galaxy, where they present the evolution of such elements with 
respect to the metallicity, \textit{i.e.} $[{\rm X}/{\rm Fe}]$ versus $[{\rm Fe}/{\rm H}]$. The relation $[{\rm Fe}/{\rm H}]$ is 
used as a measure for the metallicity, which is defined as
\begin{equation}
\left[\frac{{\rm Fe}}{{\rm H}}\right] = \log\left[\frac{\epsilon({\rm Fe})}{\epsilon({\rm H})}\right] - \log\left[\frac{\epsilon({\rm Fe})}{\epsilon({\rm H})}\right]_\odot.
\label{Fe/H}
\end{equation}
This equation compares the logarithms of the abundances of Fe and H, $\epsilon$(Fe) and $\epsilon$(H) respectively, in the
molecular cloud under consideration (first term) and solar abundances (second term). The $[{\rm Fe}/{\rm H}]$
ratio can be interpreted as a chronometer which describes the accumulation of iron in the ISM.
\cite{Timmes:1995} covered the scale from $-3.0$ to 0.0, where $[{\rm Fe}/{\rm H}]=0$ clearly represents solar metallicity. Our model 
covers an approximate range,  from $-2.5$ to 0.0. In terms of oxygen abundance, this range corresponds to 
$6.19\leq \log({\rm O/H})+12\leq8.69$, where the first refers to $[{\rm Fe}/{\rm H}]=-2.5$ and the latter to
 solar metallicity \citep{Asplund:2009}.
Apart from their standard model to describe the elemental evolution, they made two refinements:  $[{\rm N}/{\rm Fe}]$ 
calculated with convective overshoot and $[{\rm X}/{\rm Fe}]$ calculated with a factor
of 2 in the iron yields. This leads to a total of four different sets of elemental abundances
\begin{enumerate}
 \renewcommand{\theenumi}{(\arabic{enumi})}
  \item the standard model;
  \item same as (1) but $[{\rm N}/{\rm Fe}]$ calculated with convective overshoot;
  \item $[{\rm X}/{\rm Fe}]$ calculated with a factor of 2 in the iron yields, but without convective overshoot;
  \item same as (3), but with convective overshoot.
\end{enumerate}
At solar metallicity these sets coincide. The largest difference between the sets is for nitrogen at high redshift; the 
standard model predicts a very low nitrogen abundance under these conditions. Observations suggest that  nitrogen is 
primarily produced in low-metallicity massive stars, and \cite{Timmes:1995} could only reproduce this by artificially 
enlarging the numerical parameter which treats
convective overshoot (models 2 and 4). When applying this, the nitrogen abundance increases with several orders of 
magnitude, in better agreement with the observations. In general, model (4) gives the best agreement with observational 
data for all elements and we therefore use this model to obtain the abundance of all elements throughout this study.

There are however a few exceptions. This is the case for phosphorus and chlorine. The dominant molecular forms of these 
elements show only weak lines in synthetic stellar spectra, making it difficult to determine
their abundances in dwarf or field giants. For this reason, \cite{Timmes:1995} used only their standard calculations to 
predict the chlorine and phosphorus abundances relative to iron. These calculations were used in this work
to predict the initial abundances of these elements for all models.

The program \textsc{nahoon} \citep{Wakelam:2012}, which is used to simulate the chemical evolution of a particular molecular cloud, 
requires elemental abundances to be provided with respect to hydrogen and these values are obtained using
\begin{equation}
 \left[\frac{{\rm X}}{{\rm H}}\right] = \left[\frac{{\rm X}}{{\rm Fe}}\right] + \left[\frac{{\rm Fe}}{{\rm H}}\right],
\label{[X/H]}
\end{equation}
and considering the definition
\begin{equation}
\left[\frac{{\rm X}}{{\rm H}}\right] = \log\left[\frac{\varepsilon({\rm X})}{\varepsilon({\rm H})}\right] - \log\left[\frac{\varepsilon({\rm X})}{\varepsilon({\rm H})}\right]_\odot,
\end{equation}
where $\varepsilon({\rm X})$ is the abundance of certain element. Solar abundances are taken from
\cite{Asplund:2009} and are reproduced in Table \ref{tabAsplund}. 

\cite{Timmes:1995} determined the total elemental abundance for each species. Clearly, some species are depleted from 
the gas phase to form grains \citep{Graedel:1982} and the initial abundances given
to \textsc{nahoon} need to account for this. A percentage was therefore removed from the initial gas phase abundances
following \cite{Flower:2003} (see Table \ref{depletion}). The resulting initial abundances increase smoothly and 
monotonically with metallicity, which is a reflection of the elemental evolution of the Galaxy. These abundances, 
before applying depletion, are listed in Table \ref{InitialAbundancesYields}, and the electron abundances 
are set to be the sum of the initial ions abundances. All species are shown with respect to atomic H.

Apart from the atomic and molecular species, the chemical network that we will apply considers grains; so their abundances 
need to be computed as well. These can be directly determined from the depletion factors that we applied, listed in 
Table \ref{depletion}. We assume two dust compositions: carbonaceous grains (all depleted carbon) and silicate material (all other 
depleted elements), and we calculate the total mass of each composition. The resulting dust-to-gas mass ratio is in close
agreement with observational studies of the variation of this quantity with metallicity  \citep{Reshetnikov:2000,Galametz:2011}. The 
total dust volume can be derived from this by applying densities of 3.5 and 1.8 g cm$^{-3}$ for silicate and carbonaceous 
material, respectively \citep{Li:1997}. Assuming a spherical grain shape and a grain size distribution proportional 
to $n_\text{grain}(r) \propto r^{-3.5}$ \citep{Mathis:1977}, the number density of grains can be calculated. Here we assume two grain 
sizes: large grains represented by $r=0.1$~$\mu$m which cover the range from 0.05 to 0.25~$\mu$m and small grains,
ranging from 0.005 to 0.05~$\mu$m, represented by a grain with radius of $r=0.01$~$\mu$m.

\begin{table*}
\caption{Solar abundances \citep{Asplund:2009}}
 \begin{center}
  \begin{tabular}{|c|r@{.}l|}
  \hline
  X & \multicolumn{2}{c}{$\log\epsilon$(X)$_\odot$} \\
  \hline
  H  & 12&00 \\
  He & 10&93 \\
  C  &  8&43 \\
  N  &  7&83 \\
  O  &  8&69 \\
  F  &  4&56 \\
  Na &  6&24 \\
  Mg &  7&60 \\
  Si &  7&51 \\
  P  &  5&41 \\
  S  &  7&12 \\
  Cl &  5&50 \\
  Fe &  7&50 \\
  \hline
  \end{tabular} 
 \label{tabAsplund}
 \end{center}
\end{table*}

\begin{table*}
 \caption{Initial elemental abundances with respect to H nuclei for model (4).$^a$}             
 \label{InitialAbundancesYields}      
 \centering          
  \begin{tabular}{c c c c c c c }      
  \hline       
                      
  Species & \multicolumn{6}{c}{$[{\rm Fe}/{\rm H}]$}\\ 
  \hline
                           &   -2.5    &   -2.0    &   -1.5    &   -1.0    &   -0.5    &   0.0     \\ 
  \hline   
   H                       & 1.00(+00) & 1.00(+00) & 1.00(+00) & 1.00(+00) & 1.00(+00) & 1.00(+00) \\
   He                      & 0.08(+00) & 0.08(+00) & 0.08(+00) & 0.08(+00) & 0.08(+00) & 0.08(+00) \\
   N                       & 9.21(-08) & 2.05(-07) & 4.97(-07) & 1.53(-06) & 1.50(-05) & 5.62(-05) \\
   O                       & 6.12(-06) & 1.50(-05) & 3.94(-05) & 1.11(-04) & 3.07(-04) & 5.84(-04) \\
   F                       & 2.70(-10) & 5.55(-10) & 1.46(-09) & 6.16(-09) & 2.50(-08) & 6.81(-08) \\
   C$^{+}$                 & 1.22(-06) & 3.34(-06) & 1.09(-05) & 4.70(-05) & 1.82(-04) & 3.39(-04) \\
   Cl$^{+}$                & 5.30(-10) & 2.49(-09) & 7.30(-09) & 2.33(-08) & 1.02(-07) & 4.71(-07) \\
   Mg$^{+}$                & 3.47(-07) & 7.96(-07) & 2.05(-06) & 5.99(-06) & 1.75(-05) & 3.67(-05) \\
   Na$^{+}$                & 3.03(-09) & 7.16(-09) & 2.13(-08) & 9.59(-08) & 5.85(-07) & 1.83(-06) \\
   P$^{+}$                 & 3.91(-10) & 1.46(-09) & 4.03(-09) & 1.31(-08) & 6.76(-08) & 3.51(-07) \\
   S$^{+}$                 & 1.10(-07) & 3.32(-07) & 8.79(-07) & 2.38(-06) & 7.18(-06) & 1.67(-05) \\
   Si$^{+}$                & 3.73(-08) & 1.30(-07) & 4.88(-07) & 1.79(-06) & 5.99(-06) & 2.60(-05) \\
   Fe$^{+}$                & 1.00(-07) & 3.16(-07) & 1.00(-06) & 3.16(-06) & 1.00(-05) & 3.16(-05) \\
   e                       & 1.82(-06) & 4.92(-06) & 1.54(-05) & 6.04(-05) & 2.23(-04) & 4.52(-04) \\
   Small grains            & 1.85(-12) & 4.81(-12) & 1.39(-11) & 4.70(-11) & 1.55(-10) & 3.41(-10) \\
   Large grains            & 5.78(-15) & 1.50(-14) & 4.33(-14) & 1.46(-13) & 4.85(-13) & 1.06(-12) \\
   C$^{+}$/O               & 1.99(-01) & 2.22(-01) & 2.77(-01) & 4.23(-01) & 5.93(-01) & 5.80(-01) \\
   Dust to gas mass ratio  & 5.40(-05) & 1.39(-04) & 3.93(-04) & 1.26(-03) & 3.99(-03) & 8.77(-03) \\
   Grain density$^{b}$     & 3.03(+00) & 3.01(+00) & 2.97(+00) & 2.79(+00) & 2.67(+00) & 2.77(+00) \\
  \hline                 
  \end{tabular} \\
  $^a$\footnotesize{The notation $\alpha$($\beta$) stands for $\alpha$ $\times$ 10$^{\beta}$.} \\
  $^b$\footnotesize{Grain density is given in g cm$^{-3}$.}
\end{table*}

\begin{table}
 \caption{Percentage of elemental abundances remaining in the gas phase}
 \label{depletion}      
 \centering          
  \begin{tabular}{c c }     
  \hline       
                     
  Species & \multicolumn{1}{c}{Per cent in gas phase}\\ 
  \hline   
   H         & 100 \\
   He        & 100 \\
   N         & 100 \\
   O         & 70 \\
   F         & 100 \\
   C         & 40 \\
   Cl        & 50 \\
   Mg        & 0 \\
   Na        & 50 \\
   P         & 50 \\
   S         & 80 \\
   Si        & 0 \\
   Fe        & 0.04 \\
  \hline                  
  \end{tabular}
\end{table}

\subsection{Chemical model}

The \textsc{kida} chemical network kida.uva.2011\footnote{http://kida.obs.u-bordeaux1.fr/models} is used to describe the gas phase
chemistry. This network contains more than 470 atomic and molecular species connected by more than 6000 gas phase reactions. 
All reactions occur between only two reactants, since molecular clouds are not dense enough to allow three-body reactions and
are exothermic with no or small reaction barriers. In total 13 elements are included: H, He, C, N, O, Na, Mg, Cl, F, 
S, Si, P and Fe. 

\textsc{kida} was used in combination with the publicly available numerical code \textsc{nahoon} to 
create and integrate  the system of rate
equations. This code, written in \textsc{fortran90} and available through the same webpage as the \textsc{kida} chemical network, 
computes the gas phase chemistry of cold dense molecular clouds and uses the
solver \textsc{dlsode} from the \textsc{odepack} package to solve the system of equations. The chemical
evolution was computed at fixed astrophysical parameters (which is normally referred as to zero dimension model, 0D) like, for 
instance, temperature and total H density. In \textsc{nahoon}, electrons, neutral grains and grains with one electron charge are considered as
species and their abundances are also computed. The network includes neutralization reactions between positively charged 
species with negatively charged grains and reaction between electron and neutral grains (see table 3 in \cite{Wakelam:2012} for a 
complete list of reactions involving grains). The rates of these processes depend on the collision cross-section and are therefore 
different for our two grain size populations.

The grain surface reaction leading to H$_{2}$ is included in an artificial 
effective way and this is the only grain process which is accounted for.

\subsection{Physical conditions}
Dense clouds are thought to evolve from diffuse clouds. We therefore first compute the steady state
chemistry of a diffuse cloud and use the obtained abundances as initial condition for the simulations of dense clouds, with 
appropriate physical parameters. Each diffuse cloud simulation starts with all elements ionized, apart
from H, N, O, and F, and all in the atomic form.

Our choice of physical parameters is similar to the parameters used in most of gas phase astrochemical 
models. For easy comparison we chose the parameters as close as possible to the ones used 
by \cite{Wakelam:2012}, where the \textsc{nahoon} program and \textsc{kida} network was presented. For 
dense clouds, the total H density is 2$\times$10$^{4}$ cm$^{-3}$,
temperature $T$ is 10 K, visual extinction $A_{\rm V}$ is 30 mag and the cosmic ray (CR) ionization 
rate $\zeta$ is chosen to be 1.3$\times$10$^{-17}$s$^{-1}$. For clouds of such high visual extinction 
CR-induced photons are the dominant radiation source and similar results are expected using other 
high values for $A_{\rm V}$. The dependence on $\zeta$ will be discussed in more detail in 
Section~\ref{CR}. The physical parameters for both diffuse and dense clouds are summarized in
Table \ref{Physicalparameters}. Unless explicitly stated, all simulations were run with this set of 
physical parameters. The abundance of He does not vary with respect to the metallicity. All simulations 
were performed for 10$^{8}$ yr, when steady state is expected to be already reached.

\begin{table}
 \caption{Physical parameters}            
 \label{Physicalparameters}     
  \centering                        
  \begin{tabular}{c c c c}       
   \hline\hline                
   Parameter & Diffuse cloud & Dense cloud \\    
   \hline                        
     $T$ & 100 $\mathrm{K}$ & 10 $\mathrm{K}$     \\
     $n_{\rm H}$ & 1$\times10^{3}$ $\mathrm{cm^{-3}}$  & 2$\times10^{4}$ $\mathrm{cm^{-3}}$\\
     $\zeta$ & 1.3$\times10^{-16}$ $\mathrm{s^{-1}}$ & 1.3$\times10^{-17}$ $\mathrm{s^{-1}}$ \\
     $A_{\rm V}$ & 1 mag & 30 mag \\
   \hline                                  
  \end{tabular}
\end{table}

\section{Results and discussion}
\label{results}
\subsection{Influence of elemental model}
\begin{figure*}
     \centering
     \includegraphics[width=14cm]{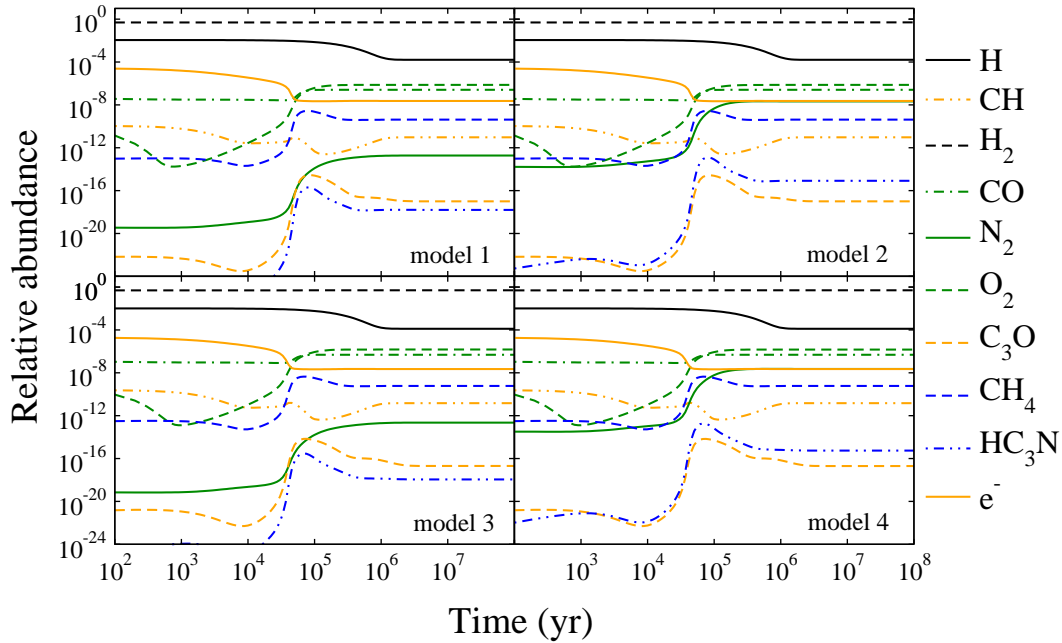}
      \caption{The chemical evolution of a dense molecular cloud for $[{\rm Fe}/{\rm H}]=-2.5$. The initial elemental abundances
      are determined from four different models taken from \protect\cite{Timmes:1995}. }
        \label{EleModels}
   \end{figure*}

As explained in Section 2.1, different sets of initial conditions could be obtained from the calculations by 
\cite{Timmes:1995}. Figure \ref{EleModels} plots the chemical evolution of a number of selected species for these four models. Additional 
depletion of the elements according to Table~\ref{depletion} has been applied. The metallicity is $[{\rm Fe}/{\rm H}]=-2.5$ in all
cases. Since the models give similar elemental abundances at solar metallicity and deviation is largest for $[{\rm Fe}/{\rm H}]=-2.5$,
these results should show the strongest effect. The figures show very similar behaviour for all species, except for N$_2$ and 
HC$_3$N. The abundance of nitrogen-bearing species is much higher in the models with convective overshoot 
for $[{\rm N}/{\rm Fe}]$ (models 2 and 4), since the amount of available nitrogen is several orders of magnitude higher.

The results presented in the remainder of this section apply model (4) to obtain the initial elemental abundances with 
additional  depletion due to the formation of grains. The percentages of elements remaining in the gas phase are given in Table
\ref{depletion}. All simulations were run using the physical parameters summarized in Table \ref{Physicalparameters}.

\begin{figure*}
     \centering
     \includegraphics[width=14cm]{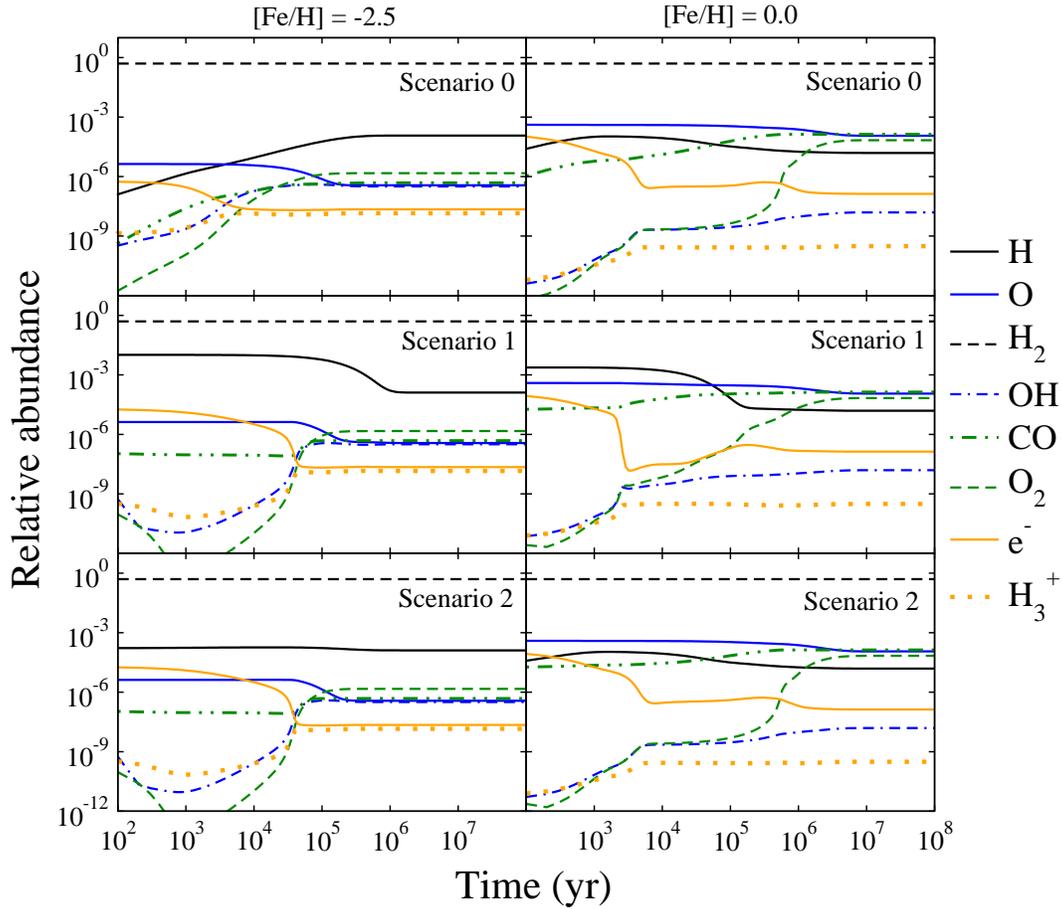}
      \caption{Evolution of important species for the gas phase chemistry according to three different scenarios (see text). Left-hand panels 
      show results for low metallicity whereas right-hand panels show results for high metallicity.}
        \label{initial_scenarios}
   \end{figure*}

Given a set of elemental abundances, different scenarios can be for the initial conditions in which these elements exist (ionized, 
atomic, or molecular form). Here we introduce three possible scenarios and chose one of them to present the results. 
Figure~\ref{initial_scenarios} shows the evolution of important species for two extreme metallicities according to the following 
scenarios: scenario 0 represents conditions typically used when simulating a dense cloud: all species ionized and hydrogen completely 
in molecular form; in scenario 1, we first simulate the evolution of diffuse clouds and use the steady state abundances as initial 
condition for dense clouds; scenario 2 is similar to scenario 1 but here the final abundances of H and H$_{2}$ obtained from the dense 
cloud simulation are taken as initial abundances for another round of dense cloud simulation. The latter scenario is 
normally chosen when one wants to study the effect of the H abundance on chemical evolution.  The steady state is more or 
less the same for all cases, but it is reached in different ways. For solar metallicity, the absence of H atoms (scenarios 0 and 2) 
pushes the chemical evolution to later times. For $[{\rm Fe}/{\rm H}]=-2.5$ scenario 0 is very different from the other scenarios. The 
results presented in the remainder of the paper are based on scenario 1.

\subsection{Metallicity dependence}
\label{Metdep}
\begin{figure*}
    \includegraphics[width=14cm]{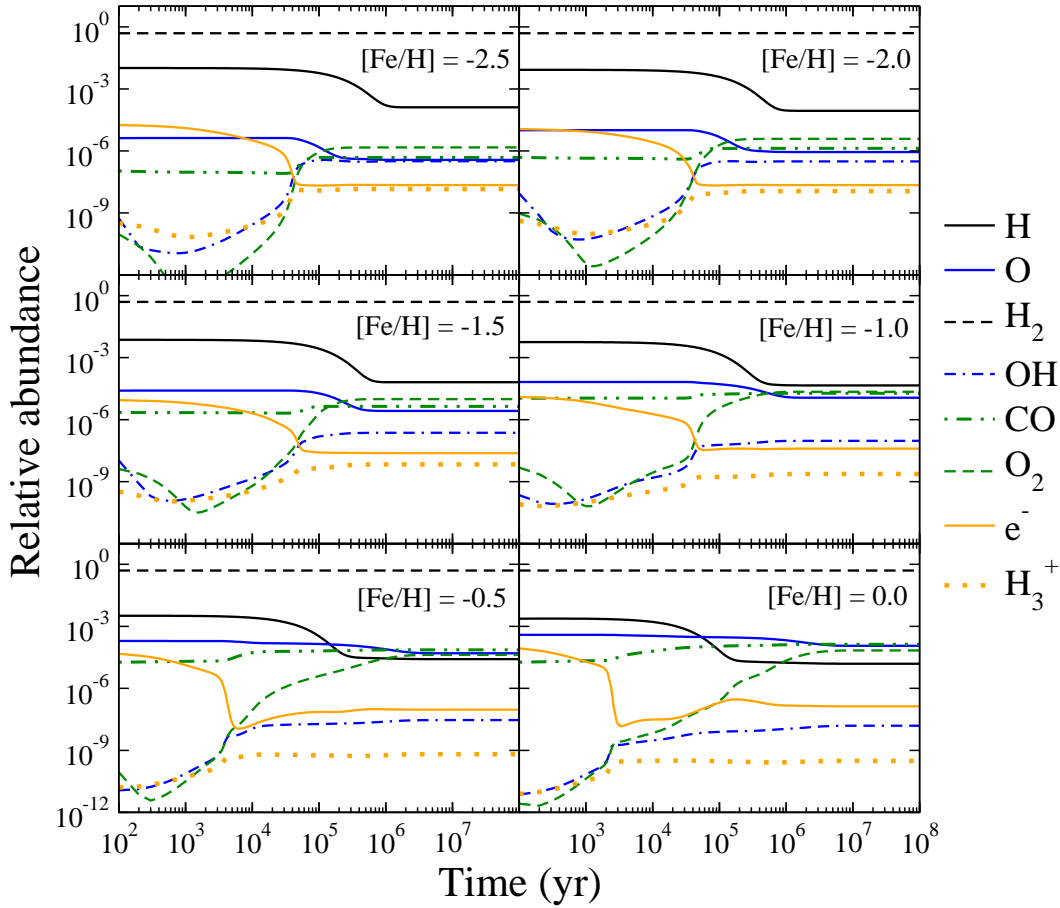}
     \caption{Simulation of the chemical evolution of a molecular cloud for different initial elemental compositions,
represented by $[{\rm Fe}/{\rm H}]$. Results are shown for species that are crucial in the chemical network.}
       \label{ChemicalEvolutionImp}
  \end{figure*}

The chemical evolution of a few species that are crucial in the chemical network is shown in 
Fig.~\ref{ChemicalEvolutionImp}. Clear differences in evolution can be observed as a function of metallicity. 
An obvious difference is the level of CO that is formed. This is a direct consequence of the amount of C and O 
available as a function of metallicity. The other oxygen-bearing molecules in Fig.~\ref{ChemicalEvolutionImp} reach steady state 
much faster for low metallicity as compared to solar metallicity. This is mainly because of the 
abundance of H$_3^+$. This species kick starts the oxygen chemistry by
\begin{equation}
 \rm H_3^+ + O \rightarrow H_2O^+ + H.
\end{equation}
From H$_2$O$^+$ a chain of reactions follows that lead to OH and ultimately to O$_2$. Since more H$_3^+$ is available at 
low metallicity, as can be seen in Fig.~\ref{ChemicalEvolutionImp} as well, O atoms are much more efficiently converted 
to O$_2$ at low metallicities. The main reason that H$_3^+$ decreases with metallicity is the increase in abundance 
of CO. CO is the main destructor of H$_3^+$ at high metallicities, whereas  for low metallicities
the CO abundance is so low that the reaction OH + H$_3^+$ becomes the main destruction pathway for H$_3^+$.

Other important species in the gas phase reaction network are electrons. The initial electron density increases with 
metallicity, since the initial abundance of free electrons is determined by the ionization of metals. So for higher 
metallicity, more electrons will initially be available. At early times, these electrons are mainly absorbed by H$_3^+$ 
and by ionized elements. Since at low metallicity the abundance of H$_3^+$ is high, this becomes an important reaction. At high
metallicities, less H$_3^+$ is available, while the initial abundance of ionized elements is higher. Therefore electrons react 
mainly with charged species. At later times, oxygen reacts efficiently with carbon chain species, destroying them and leading 
to more free electrons.

\begin{figure*}
    \includegraphics[width=14cm]{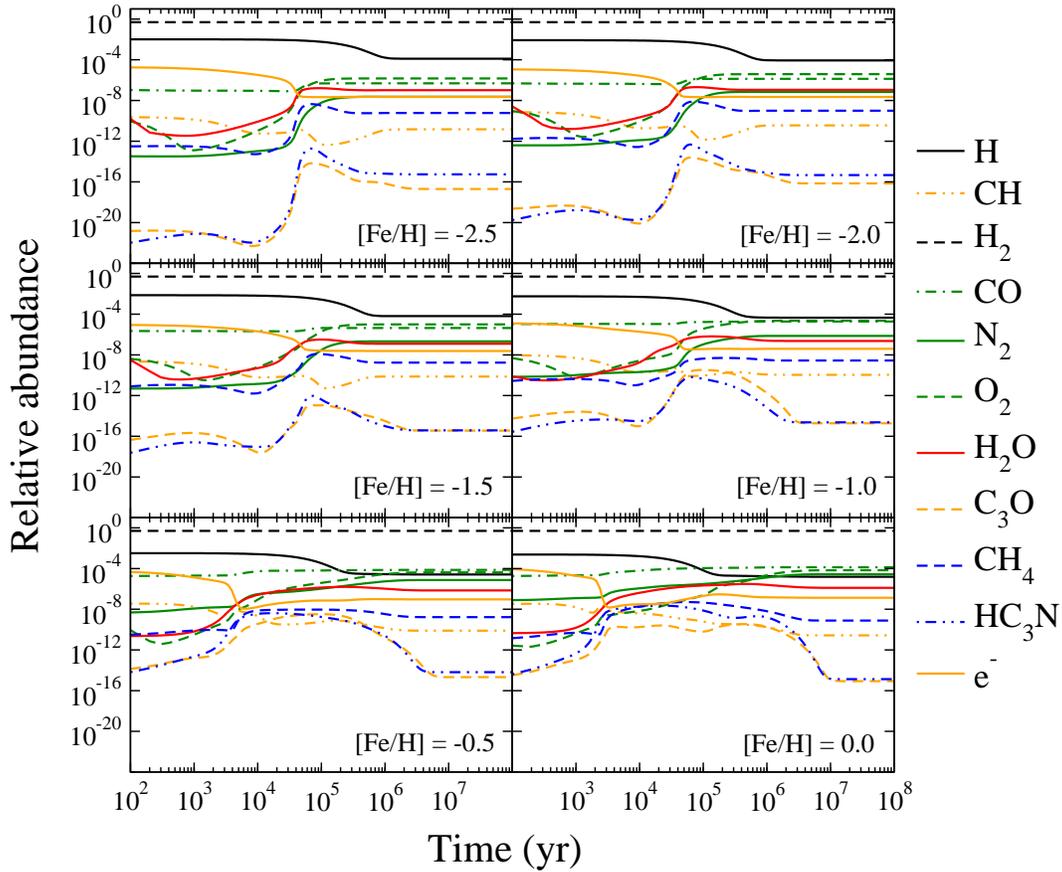}
     \caption{Simulation of the chemical evolution of a molecular cloud for different initial elemental compositions,
represented by $[{\rm Fe}/{\rm H}]$. Results are shown for simple molecules, which are representative of
the gas phase chemistry.}
       \label{ChemicalEvolution}
  \end{figure*}

Figure~\ref{ChemicalEvolution} shows the gas phase abundance of a few species. One
can see that steady state is reached later in the molecular cloud's life time as the metallicity increases. CO, for instance, 
has reached its constant final abundance as early as 1 $\times$ $10^5$~yr for $[{\rm Fe}/{\rm H}]=-2.5$ whereas 5 $\times 10^5$~yrs are
required at $[{\rm Fe}/{\rm H}]=0$. For N$_2$ the steady state time-scales are 2$\times 10^5$ and 3$\times 10^6$~yrs, respectively.
It is unclear
what the exact lifetime of a cloud is, especially at higher redshift, but it is usually assumed to be $10^{5}-10^{6}$ yr,
after which the cloud becomes dynamically unstable. If this is the case, the cloud has already reached steady state for low
metallicity conditions whereas it may not have done so for solar conditions. 

\begin{figure}
    \includegraphics[width=8.5cm]{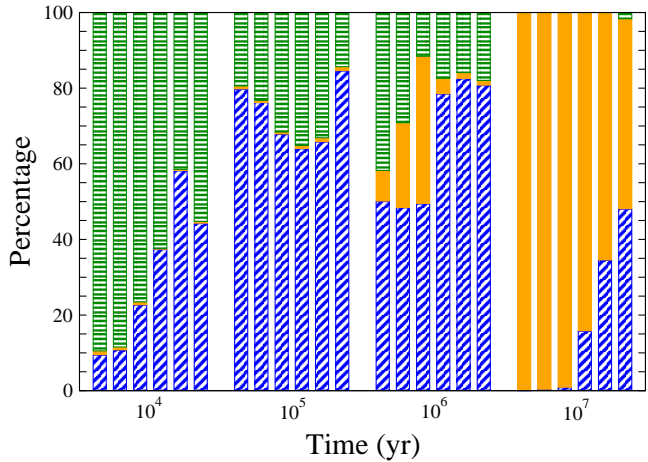}
    \caption{Percentage of molecules being destroyed (blue, oblique line), formed (green, horizontal line), or 
    in steady state (gold, solid) at different moments 
    of the molecular clouds life time for six different metallicities: from left to right ranging from $[{\rm Fe}/{\rm H}]=-2.5$ to 0.0.
    Species are considered in steady state if their relative slope satisfies $ -10^{-8} < \frac{{\rm d}n({\rm X})}{{\rm d}t}/n({\rm X}) < 10^{-8}$.}
       \label{Percentage}
  \end{figure}

Species can be classified as being formed, destroyed, or in steady state on the basis of their relative 
slopes [$\frac{{\rm d}n({\rm X})}{{\rm d}t}/n({\rm X})$]. Visual inspection of the relative slopes shows that steady state 
is reached when this parameter falls within $-1\times10^{-8}$ and $1\times10^{-8}$~yr$^{-1}$. Figure~\ref{Percentage} shows this classification in terms of 
percentage for four times and all six metallicities. Generally, three regimes can be distinguished: species are first 
predominantly formed, some reach a maximum abundance and later become predominantly destroyed and finally in steady state. 
We see that the second and third regimes are reached earlier for low metallicity than for high metallicity. This can be explained 
by the abundances of H$_3^+$ and O. At low metallicity, the high H$_3^+$ abundance leads to a fast formation of species; an example 
of this is the oxygen chemistry discussed earlier. At high metallicity, the high O abundance aids in the destruction of species 
at late times, which makes the destruction process longer. Steady state is already completely reached at 10$^{8}$ yr for all
metallicities.

\subsection{Oxygen chemistry}
\label{oxygen}
This section discusses the behaviour of oxygen-containing molecules as a function of redshift. The two most important 
reservoirs for elemental oxygen in molecular clouds are water and carbon monoxide. Water is predominantly formed through 
grain surface chemistry, which is not included in our model and the main reservoir of oxygen will therefore be CO in our 
models. We will return to the formation of H$_2$O at the end of this section.

\begin{figure}
     \centering
     \includegraphics[width=8.5cm]{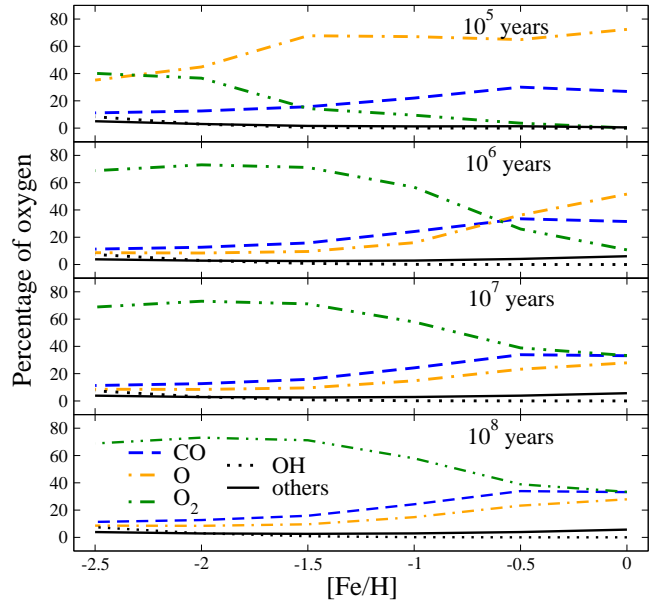}
      \caption{Contribution of different oxygen-bearing species to the total oxygen elemental abundance as a function 
      of $[{\rm Fe}/{\rm H}]$ for four different times. Only the most abundant species that together account for at least 90\% of
      oxygen are shown.}
        \label{PercentageOxygen}
   \end{figure}

\begin{figure}
     \centering
     \includegraphics[width=8.5cm]{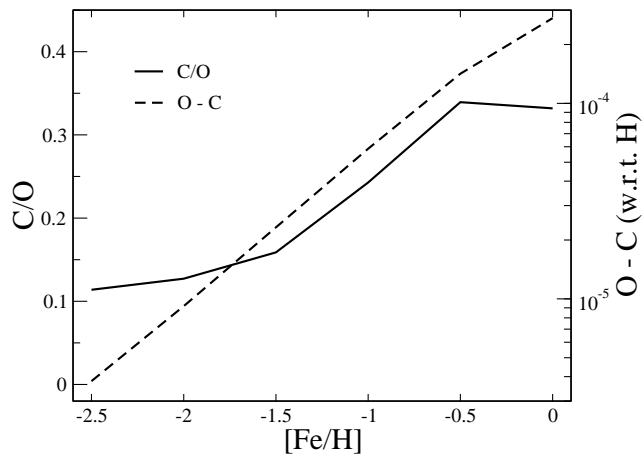}
      \caption{C/O ratio \emph{versus} $[{\rm Fe}/{\rm H}]$ (solid line) and the difference between O and C initial abundances (dashed line).
      Data are shown after depletion of the elements.}
        \label{C_O_vs_metallicity}
   \end{figure}

Figure~\ref{PercentageOxygen} plots the contribution of the different species to the total oxygen budget. This is done 
for 24 different situations: for all six metallicities at 10$^5$, 10$^6$, 10$^7$, and 10$^8$ yr. Our chemical network involves 
97 oxygen-bearing species both neutral and ionized ones. Only four species are needed to account for at least 
90\% of all available oxygen in all 24 situations: CO, O, OH, and O$_2$. In the previous section we have already discussed part 
of the oxygen chemistry and how H$_3^+$ starts the reactions to convert O to OH and finally O$_2$. This can also be seen again in 
Figure~\ref{PercentageOxygen}. 
Molecular oxygen holds around 40\% of all elemental oxygen in molecular clouds with low metallicity. This percentage 
increases with time. However, this behaviour changes as the metallicity increases. In environments of solar metallicities 
most of the O is in the atomic form, for early times, while for late times, the oxygen is almost equally shared between the three 
main O-containing species: O, O$_{2}$ and CO. The contribution of CO to the overall oxygen budget increases with metallicity. This is 
mainly because of the increase in C/O ratio. Figure~\ref{C_O_vs_metallicity} shows how the C/O
ratio increases with metallicity, but always stays below unity. This means the amount of elemental carbon is the determining 
factor in maximum amount of CO which is formed.  The CO contribution to the oxygen budget at late times follows the 
C/O ratio closely, indicating that indeed nearly all carbon is contained in CO.

On the basis of our models, one might suggest that the gas phase abundance ratio of O$_2$/CO as a function of metallicity 
would be a good observational diagnostic to test the validity of our models. Unfortunately, cold O$_2$ is very hard to 
observe since it has no permanent dipole, thus it is impossible to use O$_2$/CO as a test of our model.
   
Molecular and elemental oxygen  are both precursors for the formation of water ice through grain surface reactions. Water 
can form by different surface mechanisms as was postulated by \cite{Tielens:1982}: through the hydrogenation of O, O$_2$, 
and O$_3$. Recent experimentally studies have shown that these three routes can indeed lead to the formation of 
H$_2$O \citep{Ioppolo:2008,Miyauchi:2008,Cuppen:2010B,Dulieu:2010,Romanzin:2011}. Hydrogenation of O$_2$ and O$_3$ have 
H$_2$O$_2$ and OH as intermediates whereas O hydrogenation only has OH as intermediate. OH can react further with H or 
H$_2$ to form H$_2$O \citep{Oba:2012}. The exact route will determine the amount of formed H$_2$O$_2$, which was recently 
detected in $\rho$~Oph~A \citep{Bergman:2011} as well as the deuterium fractionation and therefore the HDO/H$_2$O ratio. The 
freeze-out of O$_{2}$ on to grains can also explain the low detection of gas phase O$_{2}$ in $\rho$~Oph~A \citep{Liseau:2012}. 
Although grain surface chemistry is not included in our models, we can speculate on the formation route based on the gas phase 
composition, since the reactants for surface reactions reach the grain surface by impinging from the gas phase \citep{Cuppen:2007A}. 
Since at low metallicity, oxygen is predominantly present in the form of O$_2$, the O$_2$ hydrogenation route is the most likely 
reaction channel leading to the formation of water ice, especially since the abundance of atomic hydrogen is relatively 
high because of the low grain density. At solar metallicity, the formation route of O + H becomes more likely, since here 
O atoms dominate. In the intermediate regime, O$_3$ hydrogenation could be more important, since here O and O$_2$ can react 
to form O$_3$. Moreover, the atomic hydrogen abundance is also lower, allowing the oxygen species to react before new H 
atoms land on the surface.

\subsection{Nitrogen chemistry}
\begin{figure}
     \centering
     \includegraphics[width=8.5cm]{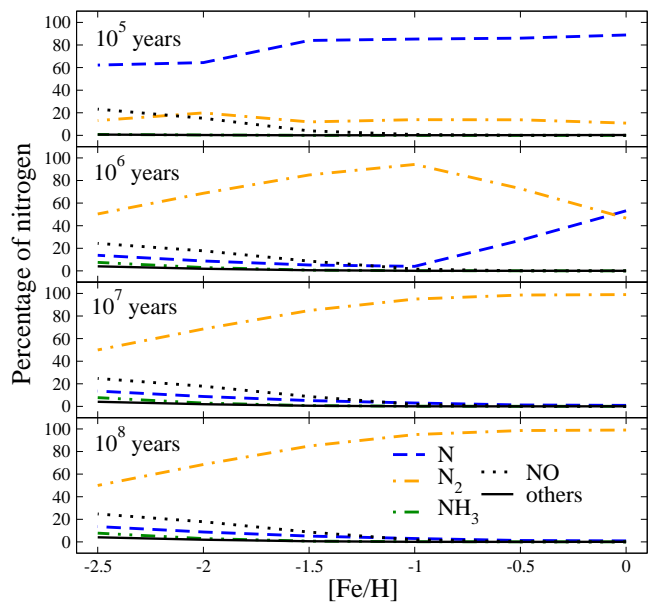}
      \caption{Similar to Fig.~\ref{PercentageOxygen} but for the nitrogen budget.}
        \label{PercentageNitrogen}
   \end{figure}

Another element of interest is nitrogen. Nitrogen is an important element in many biomolecules. Its chemistry might give 
hints whether it will be chemically available for the formation of more complex molecules. N$_2$ is for instance fairly 
inert, whereas NO is much more reactive; NH$_3$ is somewhere in between. Our network contains 123 N-bearing species. 
Figure~\ref{PercentageNitrogen} shows the contribution of the species to the total nitrogen budget. Again only a handful 
species is needed to account for a minimum of 90\% of the total nitrogen; in this case four: N, N$_2$, NO, and NH$_3$. In 
most case, however, only one or two species (N and N$_2$) are already enough to account for almost the totality of the 
nitrogen atoms. NO and NH$_3$ are mostly important at low metallicities, especially for early times. 
Figure~\ref{PercentageNitrogen} shows the elemental evolution of the Galaxy along the x-axis, while the
time evolution can be extracted by comparing the four panels top to bottom. If we analyses the top panel, which
represents the moment when it is believed that the molecular clouds start the process of collapse (10$^5$ yr), it is 
clear that atomic
nitrogen is the main holder of nitrogen, for all metallicities. This is converted to molecular nitrogen in time (second 
and third panel). This conversion is slower at high metallicity than at low metallicity. Again the difference in oxygen 
chemistry plays a crucial role, since the main reaction route to N$_2$ is
\begin{equation}
 \rm N + OH \rightarrow H + NO
\end{equation}
followed by
\begin{equation}
 \rm NO + N \rightarrow O + N_2.
\end{equation}
Since the OH abundance is high for low metallicities, this conversion will proceed faster for low metallicity.
For high metallicities, nearly all nitrogen is eventually converted to N$_2$ whereas at low metallicities a substantial 
fraction is in the form of NO, N, and NH$_3$. The nitrogen chemistry completely changes as a function of metallicity. 

At $[{\rm Fe}/{\rm H}]=-2.5$, the most dominant process which involves NO at late times is its conversion to HNO$^+$:
\begin{equation}
 \rm NO + H_3^+ \rightarrow HNO^+ + H_2
\end{equation}
 and back
\begin{equation}
 \rm HNO^+ + e^- \rightarrow H + NO,
\end{equation}
whereas this is completely unimportant at $[{\rm Fe}/{\rm H}]=0$.0. Here,
\begin{equation}
 \rm O+HNO \rightarrow NO + OH
\end{equation}
is an important channel to form NO, which is highly improbable at $[{\rm Fe}/{\rm H}]=-2.5$. At solar metallicity, many of the
common N-chemical reactions involve phosphorus species or carbon chain molecules whereas these are rather insignificant 
at $[{\rm Fe}/{\rm H}]=-2.5$.

The nitrogen-bearing species NH$_2$ can lead to the formation of NH$_3$ through
\begin{equation}
\rm  NH_2 + H_3^+ \rightarrow NH_3^+ + H_2,
\end{equation}
\begin{equation}
 \rm NH_3^+ + H_2 \rightarrow NH_4^+ + H,
\end{equation}
and
\begin{equation}
 \rm NH_4^+ + e^- \rightarrow NH_3 + H,
\end{equation}
or be destroyed by O
\begin{align}
 \rm NH_2 + O & \rm \rightarrow HNO + H\\
 & \rm \rightarrow NH + OH.
\end{align}
The first is dominant for low metallicity leading to the formation of NH$_3$ whereas the latter is dominating at high metallicity.

As explained earlier the gas phase composition can have  consequences on possible further surface 
reactions. \cite{Congiu:2012} proved experimentally that hydroxylamine (NH$_{2}$OH),
a precursor molecule in the formation of amino acids, is efficiently produced by hydrogenation of NO in interstellar ice
analogues. This process would be favoured in low metallicity environments because of the higher abundance of NO.
On the other hand, atomic nitrogen, available at mainly solar metallicities, could land on the grains and lead to the 
production of NH$_{3}$ through grain surface hydrogen atoms.

\subsection{Carbon chemistry}
\begin{figure*}
    \centering
    \includegraphics[width=16cm]{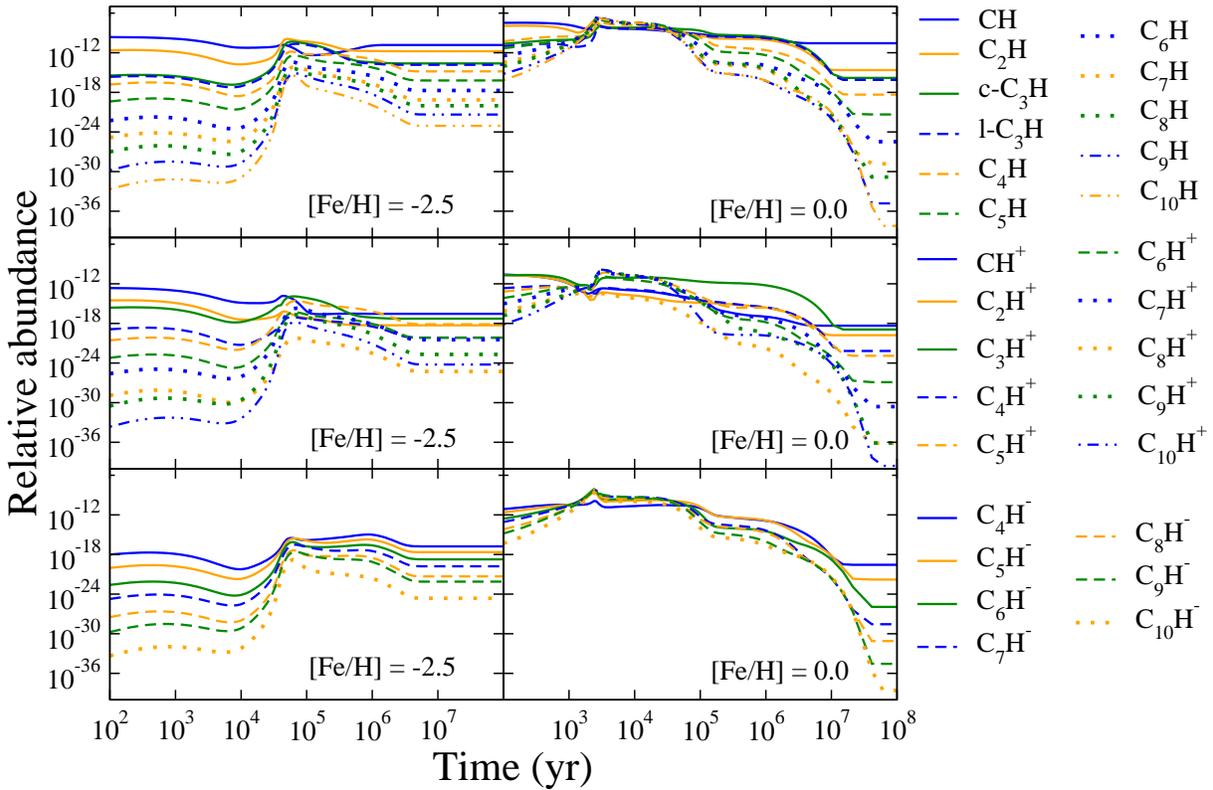}
    \caption{Evolution of carbon-bearing species with general formula C$_n$H (top panels), C$_n$H$^+$ (middle panels) and 
    C$_n$H$^-$ (bottom panels) as a function of time for $[{\rm Fe}/{\rm H}]=-2.5$ (left-hand panels) and $[{\rm Fe}/{\rm H}]=0$.0 (right-hand panels).}
       \label{CnH}
  \end{figure*}

Since the C/O ratio always stays below unity, most of the carbon will be the form of CO. However, still some carbon is
available to form more complex species. Figure~\ref{CnH} shows the abundance of species with the general formula C$_n$H 
(top panels), C$_n$H$^+$ (middle panels) and C$_n$H$^-$ (bottom panels) at the two extreme metallicities. The 
figure shows that for $[{\rm Fe}/{\rm H}]=-2.5$ the
abundance builds up more slowly than for $[{\rm Fe}/{\rm H}]=0$.0. Probably, because less carbon is available which slows down the
chemistry. This also results in a lower peak abundance. So for the more complex species the chemistry is slower; contrary 
for the small species where the chemistry is faster for low metallicity due to the higher concentration of H$_3^+$.

For the positively charged species a weak odd-even behaviour can be observed, where the odd species (in number of C) are 
more abundant. The neutral and negatively charged species do not appear to show this dependence.
After approximately 10$^{6}$~yr, the abundance quickly drops for solar metallicity and the final abundances are 
below the $[{\rm Fe}/{\rm H}]=-2.5$ values. This is due to the destruction of neutral carbon species by atomic oxygen. For low
metallicity, most of the O atoms are locked up in O$_2$. Moreover, the elemental oxygen abundance is lower to begin with. 
For other carbon species similar trends can be observed, but these are not plotted here.

\begin{figure*}
    \centering
    \includegraphics[width=18cm]{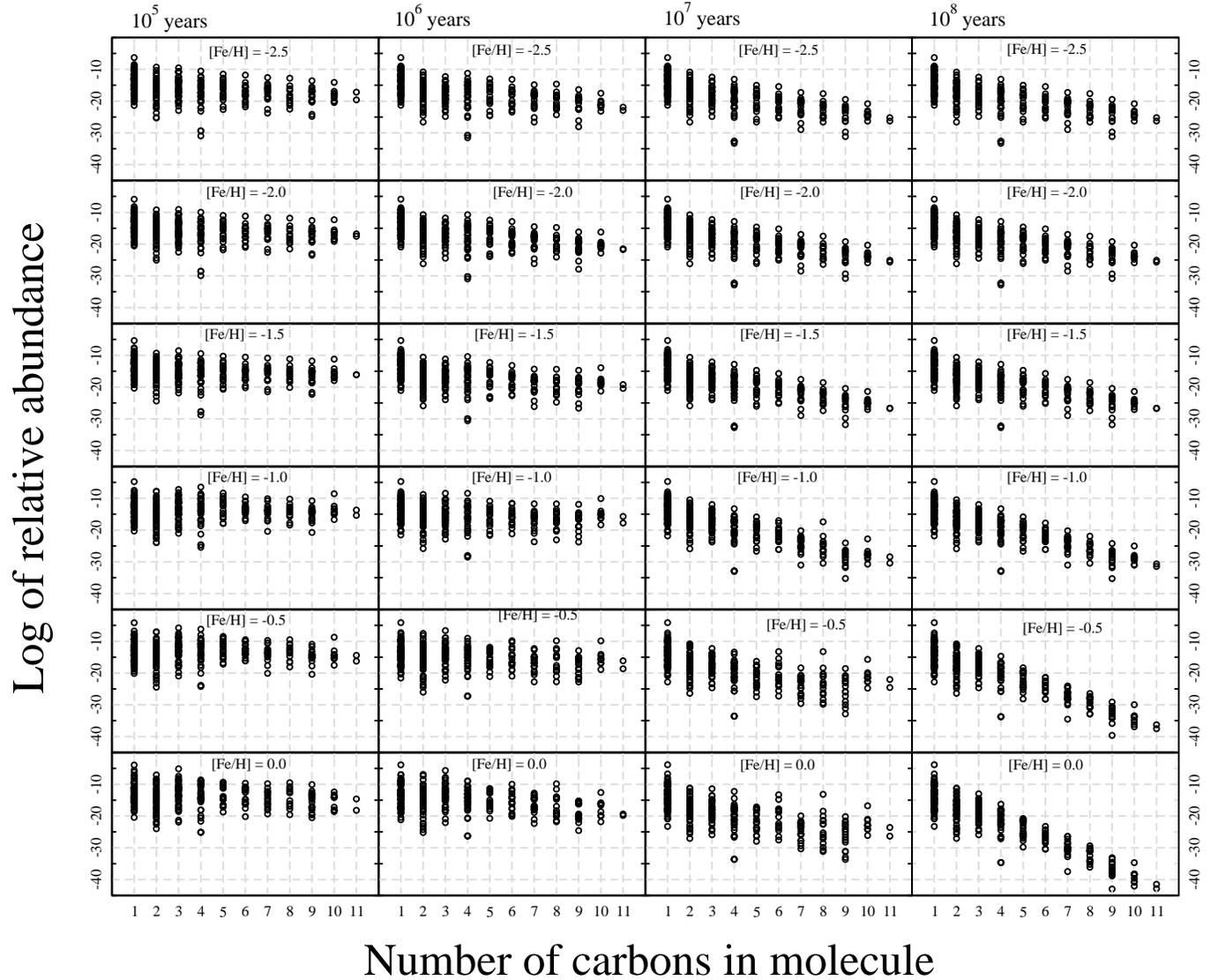}
    \caption{Effect of metallicity on carbon chain length. The abundances of carbon-bearing molecules are shown for four
different times (10$^{5}$, 10$^{6}$, 10$^{7}$, and 10$^{8}$ yr) according to the number of
carbons in the molecules.}
       \label{CChainLength}
  \end{figure*}

Figure \ref{CChainLength} plots the abundance of
carbon-bearing species versus number of carbon atoms in the molecule. This can be taken as the length of the carbon chain,
although cyclic structures like c-C$_{3}$H and C$_{6}$H$_{6}$ are included as well. This is plotted at four different 
times for each metallicity. There is a decreasing behaviour as can be expected since small molecules are the precursors for
the larger ones. 

At 10$^{5}$ yr, the plots are all more or less the same at first glance. The total amount of carbon bearing species 
however increases with metallicity, which seems logical since more carbon is available. This increase mostly occurs for
the short carbon chains. At later times, a clear drop in the large molecules can be observed, especially for high 
metallicity.
So contrary to what one might initially expect, high metallicity leads to less chemical
complexity.  This is mainly due to the high oxygen abundance.  \cite{Graedel:1982} also
observed a decrease in carbon chains as a function of the number of metals in the gas phase. This was attributed to 
destruction by electrons. Free electrons are formed through the ionization of metals. So with more metals, more electrons 
will be available
and a larger fraction of the molecular ions will be destroyed. \cite{Graedel:1982} found that better agreement with 
observations is obtained under ``low metal'' conditions. For
this reason astrochemical simulations are usually performed under these ``low metal'' conditions, where a large 
fraction of the 
gas phase metals are depleted into grains and mantles \citep{Graedel:1982,Flower:2003}. Please notice that in our simulations
we performed this depletion as well. So our high metallicity corresponds more to the ``low metallicity'' conditions applied in 
other papers, since there only one metallicity (solar) is normally considered. As mentioned in Sec.~\ref{Metdep}, ionized carbon 
chain species are mainly destroyed by oxygen at late times and high metallicity condition. This reaction leads to more free electrons, which will 
react with neutral carbon chain species. The final result is a continuous decrease in carbon chain species and increase in electron
abundance.

Similar results have been described by \cite{Millar:1990} after performing pseudo-time-dependent models to simulate 
the chemistry of dark clouds located in our Galaxy and in the Large and Small Magellanic Clouds (LMC and SMC, respectively). 
They found that the abundances of hydrocarbons vary in an unexpected way as one goes from an environment of 
higher metal abundance (Galaxy) to lower ones (LMC and SMC). It was expected that the abundances of hydrocarbons would also 
decrease in the same order, following the availability of elements. However, they found that the atomic oxygen destroys 
the hydrocarbons in accordance with our results. Moreover, they found the abundance 
of atomic O to depend on the quantity O$-$C instead of O/C, which can be seen in Fig \ref{C_O_vs_metallicity}. The lower metallicity 
$[{\rm Fe}/{\rm H}]=-2.5$ has also the lowest O$-$C value although the O/C ratio is also the highest one. This shows that there is less
atomic oxygen available to react and destroy complex species when the metallicity is lower. Figure \ref{ChemicalEvolutionImp} indeed 
shows that the abundance of atomic oxygen is lower for $[{\rm Fe}/{\rm H}]=-2.5$ during all the molecular cloud's lifetime 
compared to $[{\rm Fe}/{\rm H}]=0$.0

\subsection{Sulphur chemistry}
This section will discuss the sulphur chemistry. Although this element is present in a small number of 
species, 46 in total, still plays an important role. S-bearing molecules are often used as molecular 
clocks to predict the age of the sources where it is observed, because of the relatively fast evolution of their 
chemistry \citep{Charnley:1997B,Wakelam:2004}. It is widely accepted that in star forming regions the formation of S-bearing 
molecules is largely determined during the cold collapse phase. Atomic sulphur freezes out on grains and probably 
forms H$_2$S. After formation of the protostar, H$_2$S evaporates from the surface and starts an active chemistry in 
the warm gas ($T> 100$~K), where it reacts with H to form sulphur atoms which in turn react with the abundantly present
OH and O$_2$ to form SO and later SO$_2$ \citep[e.g.][]{Charnley:1997B}. Since this process occurs on time scales of 
roughly $10^4$ yr the relative amounts of H$_2$S, SO and SO$_2$ can be used as a chemical clock for warm gas in hot 
cores and hot corinos, but also in outflows and shocks \citep{PineauDesForets:1993, Chernin:1994, Bachiller:1997}.

\begin{figure}
     \centering
     \includegraphics[width=8.5cm]{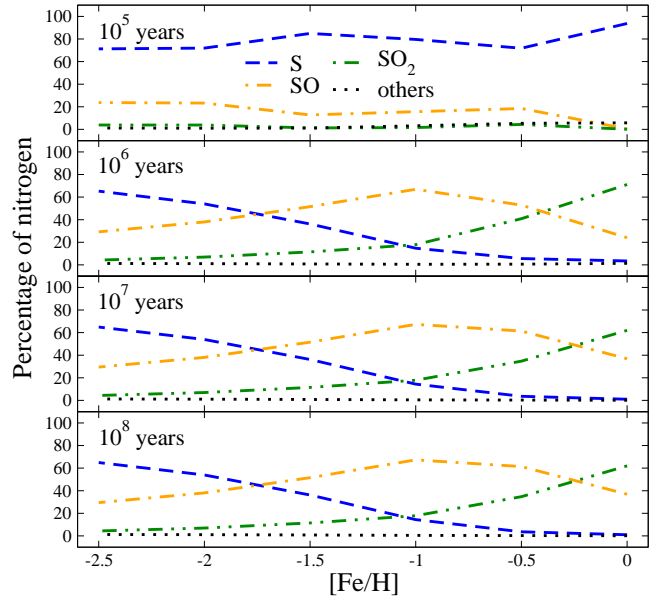}
      \caption{Similar to Fig.~\ref{PercentageOxygen} but for the sulphur budget.}
        \label{PercentageSulfur}
   \end{figure}

Since we limit our study to the cold cloud stage, it would be interesting to see how much S is available for depletion 
on to grains where it can form H$_2$S, which will kick start the sulphur chemistry at a later stage. Figure 
\ref{PercentageSulfur} shows the main sulphur-bearing species as a function of metallicity. Here three species are needed 
to account for a minimum of 90~\% of the total sulphur abundance. At early times, all sulphur is in the elemental form, 
while in late times, most of the sulphur ends locked up in SO$_{2}$, having SO as intermediary species. The amount of 
conversion increases with metallicity: for low metallicity S remains the dominant form, for solar metallicity SO$_2$ 
contains most of the sulphur, whereas at intermediate metallicity SO is more important.  This is surprising since SO 
usually forms from reaction of S with OH and O$_2$ and Figure \ref{PercentageOxygen} shows that both species are 
predominantly present at low metallicity.  
The key reactions to explain this behaviour are however the follow up reactions with SO as one of the reactants:
\begin{equation}
 \rm SO + H_3^+ \rightarrow HSO^+ + H_2
\end{equation}
and
\begin{equation}
 \rm SO + O \rightarrow SO_2.
\end{equation}
The first moves the systems down in the S~$\rightarrow$~SO~$\rightarrow$~SO$_2$ chain, whereas the latter continues the 
chain reaction. From Figure~\ref{ChemicalEvolutionImp} it can be easily understood that the first reaction dominates at 
low metallicity, whereas the latter is the main reaction route at solar metallicity, converting a large part of the sulphur 
into SO$_2$.
According to \cite{Ruffle:1999} the S depletion time scale on grains is $2.5\times10^6$ yr. Figure \ref{PercentageSulfur} 
shows that within this time frame, a significant fraction of the gas phase S has been converted to SO and SO$_2$ at high 
metallicity, which can freeze out on the grains as well. This makes the sulphur chemistry a less reliable clock for these 
conditions. At low metallicity, on the other hand, much more S is available
to form H$_2$S on the grain surfaces with the S-depletion time frame, with hardly any SO$_2$ contamination, making it a 
more clean mechanism at these conditions.

All graphs show that the majority of the species are in neutral form. This is because under these conditions the cloud is 
shielded from most ionization processes.

\subsection{Density dependence}
\begin{figure*}
     \centering
     \includegraphics[width=14cm]{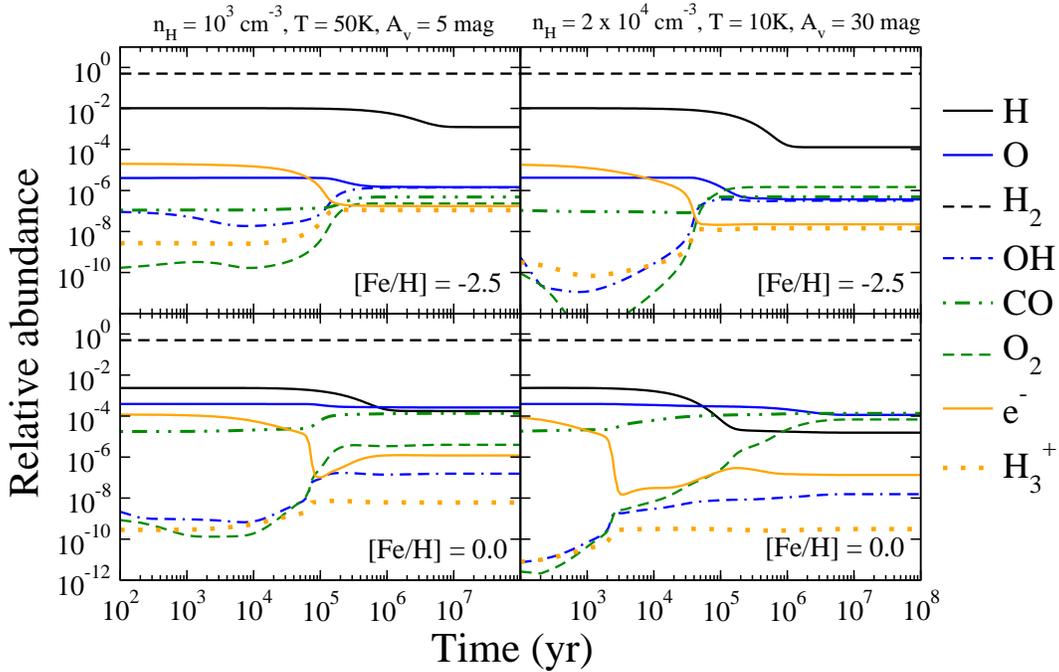}
      \caption{Comparison between lower and higher densities. Left-hand panels: time evolution of diffuse/translucent clouds
      with total hydrogen density $n_\text{H}=10^{3}$ cm$^{-3}$, temperature $T=50$~K and visual extinction $A_\text{V}=5$~mag. Right-hand panels: time
      evolution of dense clouds with the physical parameters described in Table \ref{Physicalparameters}. Top panels show results
      for low metallicities whereas high metallicity is shown on bottom panels. A CR ionization rate of
      $\zeta=1.3\times10^{-17}$ s$^{-1}$ was used in all simulations.}
        \label{density}
   \end{figure*}

Extragalactic sources are usually not spatially resolved and the observations consist of contributions by different density 
regions, mainly those regions of low densities. We therefore performed simulations of a translucent cloud, with a lower total hydrogen 
density of $n_\text{H}=10^{3}$ cm$^{-3}$, temperature $T=50$~K and visual extinction $A_\text{V}=5$~mag, to compare to the dense cloud 
results. This comparison is shown in Fig.~\ref{density}. The left-hand panels show the time evolution of the translucent cloud whereas the 
right-hand panels show the evolution of the dense cloud. In both cases the initial chemical composition of the cloud was taken from a diffuse 
cloud simulation. For the translucent cloud, the chemical time scale remains more or less the same for low metallicities whereas the 
chemical time-scale is much more constraint for high metallicity. Chemistry starts later because of the lower density, but steady state is 
reached much faster. The absolute relative abundances are also quite different going from the dense to the translucent cloud. The 
stable species O$_2$ decreases whereas all other species increase.

The search for observational evidence of molecules in external galaxies started already a few decades ago. Up to the present date,
around 60 molecules have been detected in extragalactic sources (see ``http://www.astro.uni-koeln.de/cdms/molecules" for an
updated list). Most of these molecules are detected in nearby galaxies \citep[e.g.][]{Rickard:1975,Martin:2003}.
To relate our model results to observations, we will first need to relate metallicity to redshift, which depends on the cosmological 
model applied. Figure 9 in \cite{Chiappini:1999} shows this relationship for different cosmologies. For more standard values, a 
metallicity of $[{\rm0/Fe}]=0.6$, which corresponds to $[{\rm Fe/H}]=-2.5$ according to \cite{Timmes:1995}, results in a redshift 
between 3.8 and 3. Species have been detected at such high redshift and even higher \citep[e.g.,][]{Walter:2003,Wagg:2005,Guelin:2007}, 
but fractional abundances have not been derived. The \textit{Hershel} Dwarf Galaxy Survey has observed lines like [CII], [OI], [OIII] 
and [NII] for low-metallicity galaxies \citep{Madden:2013}. Unfortunately, again no abundances have been presented so far. This
makes it difficult at the present stage to make a direct comparison between our model results and observations.

\subsection{Cosmic ray ionization dependence}
\label{CR}
CRs play a crucial role on the chemistry of the ISM since they are the main ionization mechanism capable to 
ionize many of the most abundant species and, therefore, initiating the reaction network. For example, reaction between CRs and 
H$_{2}$ produces electrons
\begin{equation}
 \rm H_2 + CR \rightarrow H_2^+ + e^- + CR',
\end{equation}
and triggers the formation of H$_{3}^{+}$ in the follow-up reaction:
\begin{equation}
 \rm H_2^+ + H_2 \rightarrow H_3^+ + H.
\end{equation}
As discussed in Section \ref{Metdep}, these are two of the most determining species for the chemical evolution of our molecular clouds. 
The outcome of our simulations will therefore strongly depend on the value chosen for the CR ionization rate. Unfortunately, this rate 
is highly uncertain. Theoretical and observational studies give a range that covers orders of magnitude: from 10$^{-18}$ to 
10$^{-15}$ s$^{-1}$; although recent studies point to a value on the order of 10$^{-16}$ s$^{-1}$ for diffuse clouds and about 0-2 
orders of magnitude lower in dense clouds \citep{Indriolo:2012}. Our initial choice of $\zeta = 1.3 \times 10^{-16}$ s$^{-1}$ for diffuse 
clouds and $\zeta = 1.3 \times 10^{-17}$ s$^{-1}$ for dense clouds is in accordance with this data.

   \begin{figure*}
     \centering
     \includegraphics[width=16cm]{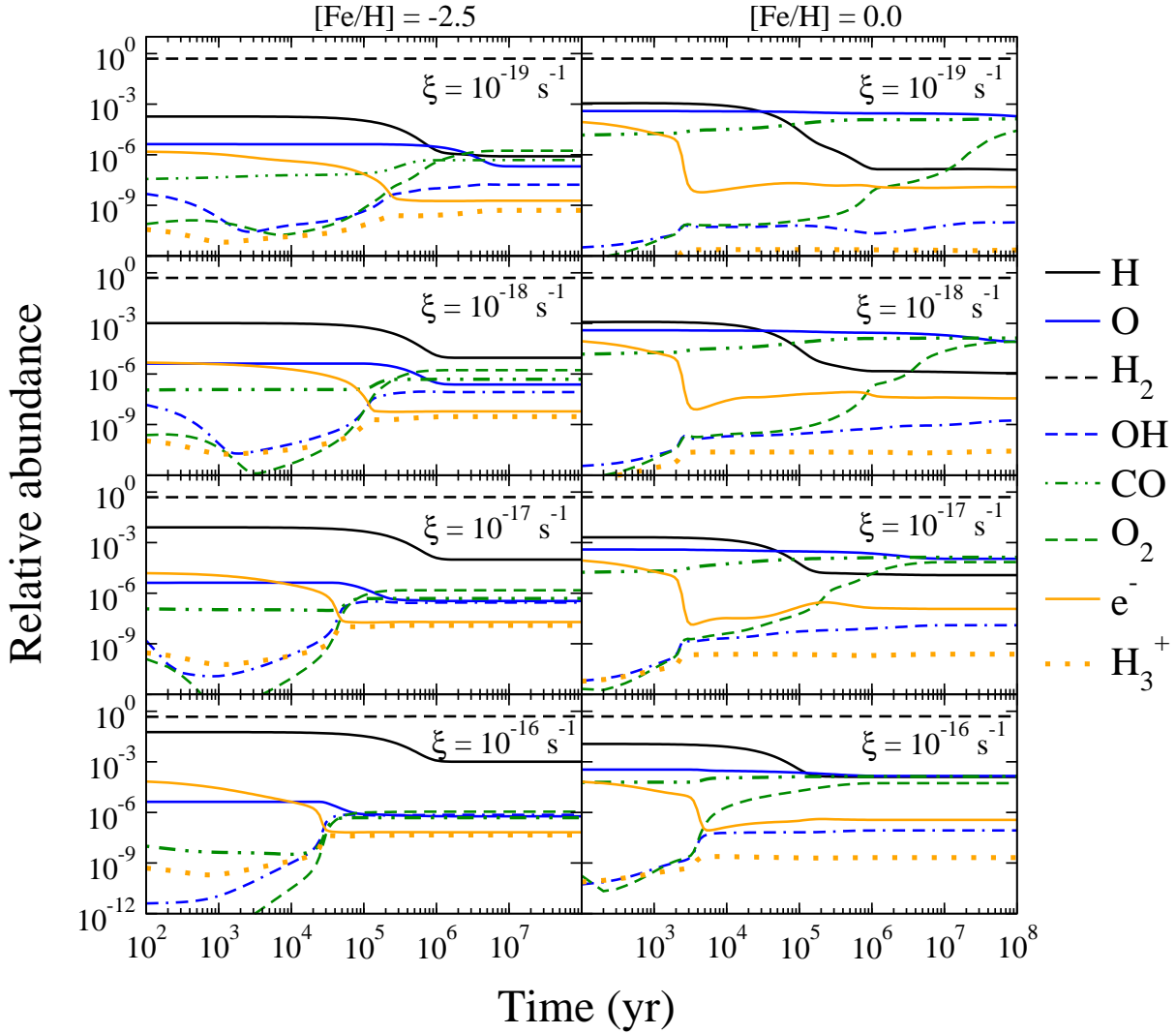}
      \caption{Dependence on the CR ionization rate in the evolution of crucial molecules in dense molecular clouds.
      Simulations were performed for the two extreme metallicities. The CR ionization rate used is noted and each simulation was 
      preceded by the simulation of a diffuse cloud with CR ionization rate of one order of magnitude higher.}
        \label{ionization}
   \end{figure*}

Figure \ref{ionization} shows the impact of different values of CRs ionization rate on the evolution of the
crucial species. Four different CR ionization rate values, ranging from $\zeta$ = 10$^{-19}$ to 10$^{-16}$ s$^{-1}$, 
were chosen to investigate the impact on the chemical evolution at the two extreme metallicities. Each simulation was performed 
using the parameters described in Table \ref{Physicalparameters} but with the CR ionization rate as noted in the 
Figure \ref{ionization}. The CR ionization rate in the preceding diffuse cloud simulations was chosen to be one order of magnitude higher.

\begin{figure}
     \centering
     \includegraphics[width=8.5cm]{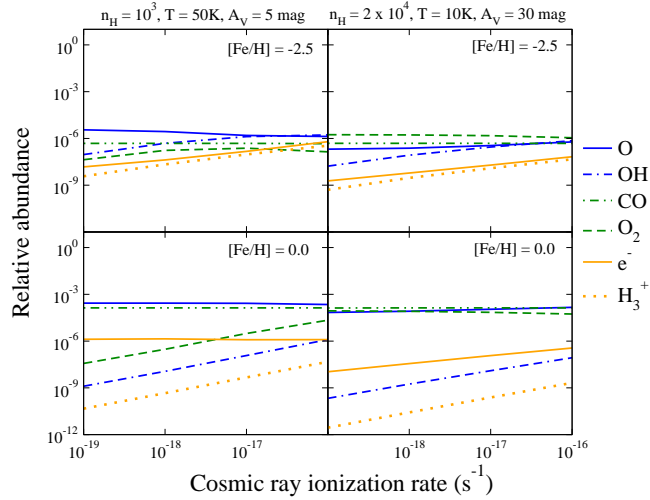}
      \caption{Dependence of the steady state abundance on the CR ionization rate for two different total hydrogen density.}
        \label{abundance_CR_1}
   \end{figure}

It is clear that the steady state of the considered molecules is reached earlier as the CR ionization
rate increases, for both metallicities. The production of e$^{-}$ and H$_{3}^{+}$ increases with the CR
ionization rate as is expected, since these species are direct products of interaction between CRs and H$_{2}$. Their steady state 
abundance together with OH follows a power-law dependence on the CR ionization rate as can be seen in Figure~\ref{abundance_CR_1}. The
abundances of O, O$_2$ and, CO appear to remain unaffected by the CR ionization rate except for high CR rates in 
translucent clouds at low metallicity. Notice that for the lowest CR ionization rates at solar metallicity steady state was reached beyond 
$10^7$~yr.

It is well established that the chemistry of dense molecular clouds may present bistability when varying physical parameters 
like the CR ionization rate and depletion level \citep[e.g.,][]{LeBourlot:1993,LeBourlot:1995,Lee:1998}. This phenomenon is suggested 
to be a mathematical consequence of the non-linearity of the rate equations that describe the chemical evolution of molecular 
clouds \citep{Shalabiea:1995}. Species involved in this phenomenon are H$_{3}^{+}$, O$_{2}$ and S$^{+}$ and the reactions connecting 
them, as described by \cite{Boger:2006}. In this section we vary both the initial elemental abundance and the CR ionization rate. We 
see no signs of bistability in Figures~\ref{ionization} and \ref{abundance_CR_1}, since here the steady
state abundances change smoothly with CR ionization rate; the O$_{2}$ steady state abundance appears even to be unaffected by the 
CR ionization rate in most cases, except for the case of low total hydrogen density and high metallicity, where it increases monotonically. 
Although no sign of bistability was found with the described simulations, this phenomenon could still be present 
here. To explore the presence of bistable solutions in detail, one needs to perform simulations with a range of random initial 
conditions. However, such performance is beyond the scope of the present work.

\section{Conclusions}
\label{conclusions}
We have studied the role of the elemental evolution of the Galaxy on the chemical evolution of cold and dense molecular 
clouds. Violent processes leading to the death of stars deliver to the ISM new elements produced 
in the interior of the stars. Such elements enrich the surrounding gas, which will evolve and form new stars and 
planetary systems. Therefore, new generations of stars are born with different metallicities. In this paper, we simulate the
chemical evolution of molecular clouds from a set of different initial elemental abundances, reflecting the change on the 
metallicity leaded by the elemental evolution of the Galaxy. Our main results are listed below:

\begin{enumerate}

 \renewcommand{\theenumi}{(\arabic{enumi})}
  \item The changes in gas phase chemistry can be explained by the changes in abundance of three species: 
  electrons, H$_3^+$, and O.
  \item Electrons mainly charge carbon chain species at high metallicity, which in turn will be destroyed by atomic oxygen.
  \item Since CO is less abundant at low $[{\rm Fe}/{\rm H}]$ because of the low availability of elemental C and O, H$_3^+$ can sustain
  higher abundances. This triggers the formation of OH and O$_2$ from atomic oxygen. It further destroys SO and PO and 
  facilitates the conversion of NH$_2$ into NH$_3$.
  \item Because of the low O abundance in general and its quick conversion into O$_2$, less O atoms are available at low 
  $[{\rm Fe}/{\rm H}]$ to destroy large carbon species, to convert SO into SO$_2$, and to prevent the formation of NH$_3$ from NH$_2$.
  \item Higher CR ionization leads to higher abundances of H$_3^+$, OH and electrons. The steady state is reached earlier with 
  higher cosmic ray ionization rate.
  \end{enumerate}
   
The new generations of ground-base interferometers, such as ALMA, will increase the
sensitivity of observations allowing unequivocal detection of molecules in extragalactic sources, at higher
redshifts, forcing a new development of astrochemical models able to take into account the new chemical complexity that 
will come up with these new facilities. In the near future, it will be possible to have a better comprehension of the 
chemistry happening in different redshift from the observational point of view, putting the current work to the test.

\section*{Acknowledgments}
EMP and HMC  acknowledge the European Research Council (ERC-2010-StG, Grant Agreement no. 259510-KISMOL) for 
financial support. HMC is grateful for support from the VIDI research program 700.10.427, which is financed by 
The Netherlands Organization for Scientific Research (NWO). EMP also thanks FAPERJ for financial support during the 
early developments of this research. The authors would like to thank Eric Herbst, Tom Millar and the anonymous referee for
interesting discussions and suggestions. 

\bibliography{bibliography}
\bibliographystyle{mn2e}

\label{lastpage}
\end{document}